\documentclass[12pt,preprint]{aastex} 
\shorttitle{DIG Halo Kinematics in NGC 891}
\shortauthors{Heald et al.}

\newcommand{\kms}[1]{#1\,\mathrm{km\,s^{-1}}}
\newcommand{\kmskpc}[1]{#1\,\mathrm{km\,s^{-1}\,kpc^{-1}}}
\newcommand{\ha}{H$\alpha$}
\newcommand{\hi}{H$\,$I}
\newcommand{\nii}{[N$\,$II]}
\newcommand{\sii}{[S$\,$II]}
\newcommand{\hii}{H$\,$II}
\newcommand{\zi}{$25\arcsec\,<\,z\,<\,45\arcsec$}
\newcommand{\zii}{$45\arcsec\,<\,z\,<\,65\arcsec$}
\newcommand{\ziii}{$65\arcsec\,<\,z\,<\,105\arcsec$}
\newcommand{\zipc}{$1.2\,\mathrm{kpc}\,<\,z\,<\,2.1\,\mathrm{kpc}$}
\newcommand{\ziipc}{$2.1\,\mathrm{kpc}\,<\,z\,<\,3.0\,\mathrm{kpc}$}
\newcommand{\ziiipc}{$3.0\,\mathrm{kpc}\,<\,z\,<\,4.8\,\mathrm{kpc}$}

\begin{document}

\title{Integral Field Unit Observations of NGC 891:\\Kinematics of the Diffuse Ionized Gas Halo}
\author{George H. Heald\altaffilmark{1} and Richard J. Rand\altaffilmark{1}}
\affil{University of New Mexico, Department of Physics and Astronomy, 800 Yale Boulevard NE, Albuquerque, NM 87131}
\email{gheald,rjr@phys.unm.edu}
\author{Robert A. Benjamin}
\affil{University of Wisconsin -- Whitewater, Department of Physics, 800 West Main Street, Whitewater, WI 53190}
\email{benjamir@uww.edu}
\and
\author{Matthew A. Bershady}
\affil{University of Wisconsin -- Madison, Department of Astronomy, 475 North Charter Street, Madison, WI 53706}
\email{mab@astro.wisc.edu}
\altaffiltext{1}{Visiting Astronomer, Kitt Peak National Observatory, National Optical Astronomy Observatory, which is operated by the Association of Universities for Research in Astronomy, Inc. (AURA) under cooperative agreement with the National Science Foundation.}
\setcounter{footnote}{1}

\begin{abstract}
We present high and moderate spectral resolution spectroscopy of diffuse ionized gas (DIG) emission in the halo of NGC 891. The data were obtained with the SparsePak integral field unit at the WIYN\footnote{The WIYN Observatory is a joint facility of the University of Wisconsin-Madison, Indiana University, Yale University, and the National Optical Astronomy Observatory.} Observatory. The wavelength coverage includes the \nii $\lambda\lambda\,6548,6583$, \ha, and \sii $\lambda\lambda\,6716,6731$ emission lines. Position-velocity (PV) diagrams, constructed using spectra extracted from four SparsePak pointings in the halo, are used to examine the kinematics of the DIG. Using two independent methods, a vertical gradient in azimuthal velocity is found to be present in the northeast quadrant of the halo, with magnitude approximately $\kmskpc{15-18}$, in agreement with results from \hi\ observations. The kinematics of the DIG suggest that this gradient begins at approximately 1 kpc above the midplane. In another part of the halo, the southeast quadrant, the kinematics are markedly different, and suggest rotation at about $\kms{175}$, much slower than the disk but with no vertical gradient. We utilize an entirely ballistic model of disk-halo flow in an attempt to reproduce the kinematics observed in the northeast quadrant. Analysis shows that the velocity gradient predicted by the ballistic model is far too shallow. Based on intensity cuts made parallel to the major axis in the ballistic model and an \ha\ image of NGC 891 from the literature, we conclude that the DIG halo is much more centrally concentrated than the model, suggesting that hydrodynamics dominate over ballistic motion in shaping the density structure of the halo. Velocity dispersion measurements along the minor axis of NGC 891 seem to indicate a lack of radial motions in the halo, but the uncertainties do not allow us to set firm limits.
\end{abstract}

\keywords{galaxies: halos --- galaxies: individual (NGC 891) --- galaxies: kinematics and dynamics}

\section{Introduction}\label{section:introduction}

\setcounter{footnote}{2}

In recent years, deep observations of external spiral galaxies have led to the realization that the multiphase nature of the ISM in disks is found in halos as well. Gaseous halos are found to contain neutral hydrogen \citep[\hi; e.g.,][]{i94,ssv97}, diffuse ionized gas \citep[DIG; e.g.,][]{rkh90,d90,rd03}, hot X-ray gas \citep[e.g.,][]{bp94,tprbd06}, and dust \citep[e.g.,][]{hs99,im06}. The origin of the multiphase gaseous halos remains unclear, but the gas is generally considered to be participating in a star formation-driven disk-halo flow, such as that described by the fountain or chimney model \citep{sf76,b80,ni89}, being accreted from companions \citep[e.g.,][]{vs05}, or originating in a continuous infall \citep{trsp02,kmwsm05}. Determining which of these pictures is dominant in halos, and thus gaining a better understanding of how disks and halos share their resources, will have a significant impact on how we view the evolution of galaxies.

Observations of extraplanar DIG (EDIG) in edge-on systems provide strong lines of evidence supporting the idea that star formation in the disk is responsible for the large quantities of gas observed in halos. Filamentary structures, often rooted in \hii\ regions in the disk, are seen in many halos \citep[e.g.,][]{r96}. Additionally, the total amount of EDIG emission correlates with a measure of the star formation rate per unit area, the surface density of far infrared luminosity ($L_{\mathrm{FIR}}/D_{25}^2$, where $D_{25}$ is the optical isophotal diameter at the 25th magnitude) \citep{r96,mv03,rd03}. \hi\ observations have revealed the presence of vertical motions in some face-on disks \citep[e.g.,][]{ks93}; these vertical motions are thought to be indicative of injection of matter into the halo.

Accretion, on the other hand, is an attractive alternative for the origins of the cold halo gas, particularly in systems displaying morphological or kinematic lopsidedness, or obvious signs of interactions \citep[see, e.g.,][]{vs05}. Increasingly deep \hi\ observations reveal clear connections between disks and companions, and suggest important connections between disks and external sources of matter \citep[see also][]{s99}. Continuous infall of halo gas over a galaxy's lifetime has been invoked to explain the star formation histories of galaxies and the ``G-dwarf problem'' \citep[e.g.,][]{p97}. It is possible that a good understanding of the kinematics of such extraplanar gas, perhaps by considering the different gas phases in parallel, may help reveal the importance of all of these processes.

To begin to understand the dynamics of gaseous halos in external spirals, edge-on galaxies are good targets because confusion between emission from the disk and from the halo is minimized. A prime target for such studies has been NGC 891, a nearby edge-on spiral. Early observations by \citet{sa79} revealed the presence of a thick vertical \hi\ distribution. Detailed three-dimensional modeling of deep Westerbork Synthesis Radio Telescope (WSRT) \hi\ data by \citet{ssv97} indicated that the halo gas lags the disk by about 25 to $\kms{100}$. Even deeper WSRT observations and more detailed modeling of the kinematics have recently been performed \citep{foss05}, showing that a gradient in azimuthal velocity with height above the midplane ($z$) exists, with magnitude $\kmskpc{15}$.

For the DIG component of halos, progress has been made only recently in robustly measuring rotational properties. Early work focused on mean velocities determined from slit spectra \citep[e.g.,][]{r00,tdsur00}, but because the distribution of emitting gas along the line of sight contributes to the shape of the line profile (this is especially important in edge-ons), mean velocities are not indicative of the rotation speeds. To properly explore the kinematics of DIG halos, high spectral resolution emission line data in two spatial dimensions are required, and can be obtained with Fabry-Perot etalons or Integral Field Units (IFUs). This method has been used by \citet[][hereafter Paper I]{hrbch06}, who measured a vertical gradient in azimuthal velocity in the halo of NGC 5775, with magnitude $\kmskpc{\approx 8}$.

The ultimate goal of these kinematic studies is to gain insight into the physics of the disk-halo interaction and the origin of halo gas. To that end, a handful of simple but tractable models, representing distinct physical pictures, have been developed in an attempt to match the observations. \citet{cbr02} constructed a purely ballistic model of a galactic fountain. Although the model naturally predicts a vertical gradient in rotational velocity, it is too shallow to match the observations of NGC 5775 (Paper I). Recently, \citet{fb06} also considered a ballistic model; the velocity gradient produced by their model is too shallow when compared to the \hi\ kinematic data of NGC 891 presented by \citet{foss05}. Both of these models treat the material participating in the disk-halo flow as a collection of non-interacting particles. Physically, this picture is appropriate if the density contrast between the orbiting clouds and the ambient medium is sufficiently high that the presence of the latter may be neglected. At the other extreme, two distinct models which treat the halo gas hydrostatically or hydrodynamically have been considered. A hydrostatic model of a rotating gaseous halo \citep{bcfs06} has been shown to reproduce the \hi\ results of \citet{foss05}. Such a model corresponds to a quiescent halo with no disk-halo interaction of the type considered in the ballistic models. The other model is a smoothed particle hydrodynamic simulation but one in which the halo gas has a completely different origin: accretion during galaxy formation \citep{kmwsm05}. This model, too, has been able to reproduce accurately the observed \hi\ kinematics of the halo of NGC 891. The models described here cover an extremely broad range of physical possibilities. When the results are considered together with the observations of gaseous halos described above, no individual physical model is completely satisfactory. Further observational and theoretical work will be necessary to help resolve this issue.

Here we present high and moderate resolution IFU observations of EDIG emission in the halo of NGC 891. Classified as Sb in the Third Reference Catalogue of Bright Galaxies \citep[RC3;][]{ddcbpf91}, NGC 891 has a systemic velocity of $\kms{528}$ (RC3), and we take the distance to be $9.5\,\mathrm{Mpc}$ after \citet{vs81}. At that distance, $1\,\mathrm{kpc}$ subtends about 22\arcsec. The surface density of far infrared luminosity ($L_{\mathrm{FIR}}/D_{25}^2$) is $3.19\,\times\,10^{40}\,\mathrm{erg\,s^{-1}\,kpc^{-2}}$ \citep{rd03}, which is indicative of moderate ongoing star formation via the prescription given by \citet{k98} for $L_{\mathrm{FIR}}$. The EDIG component has been paid great observational attention due to NGC 891's proximity, similarity to our own Milky Way, and edge-on orientation. The first studies \citep{rkh90,d90} made use of narrow-band imaging to reveal an extremely prominent DIG halo, extending up to at least $4\,\mathrm{kpc}$ above the plane, consisting of vertical filamentary structures superimposed on a smooth but asymmetric background. Previous ground-based imaging \citep[e.g.,][]{hs00}, high spatial resolution \emph{HST} imaging \citep{rdwn04}, and deep long-slit spectroscopy \citep[e.g.,][]{r97,ogr02}, have greatly enhanced our understanding of the distribution and physical conditions of the gas. For a brief discussion on the current status of how these lines of evidence fold into the larger picture of gaseous halos, see \citet{d05}.

A critical parameter for accurate determination of velocities later in the paper is the inclination angle of NGC 891. Estimates of the inclination include $i>87.5\degr$ \citep{sa79}; $i\geq 88.6\degr$ \citep{rvkgs87}; and, most recently, $i=89.8 \pm 0.5\degr$ \citep{kv05}. The last result is based on modeling of stellar kinematics and dust extinction. The inclination angle is extremely close to $i=90\degr$, and we adopt that value in this paper. Slight deviations from this value will not significantly alter our results.

This paper is arranged as follows. We describe the observations and the data reduction steps in \S \ref{section:observations}. The halo kinematics are examined in \S \ref{section:kinematics}, and the ballistic model is compared to these results in \S \ref{section:ballistic}. We conclude the paper in \S \ref{section:conclusions}.

\section{Observations and Data Reduction}\label{section:observations}

\begin{figure}
\plotone{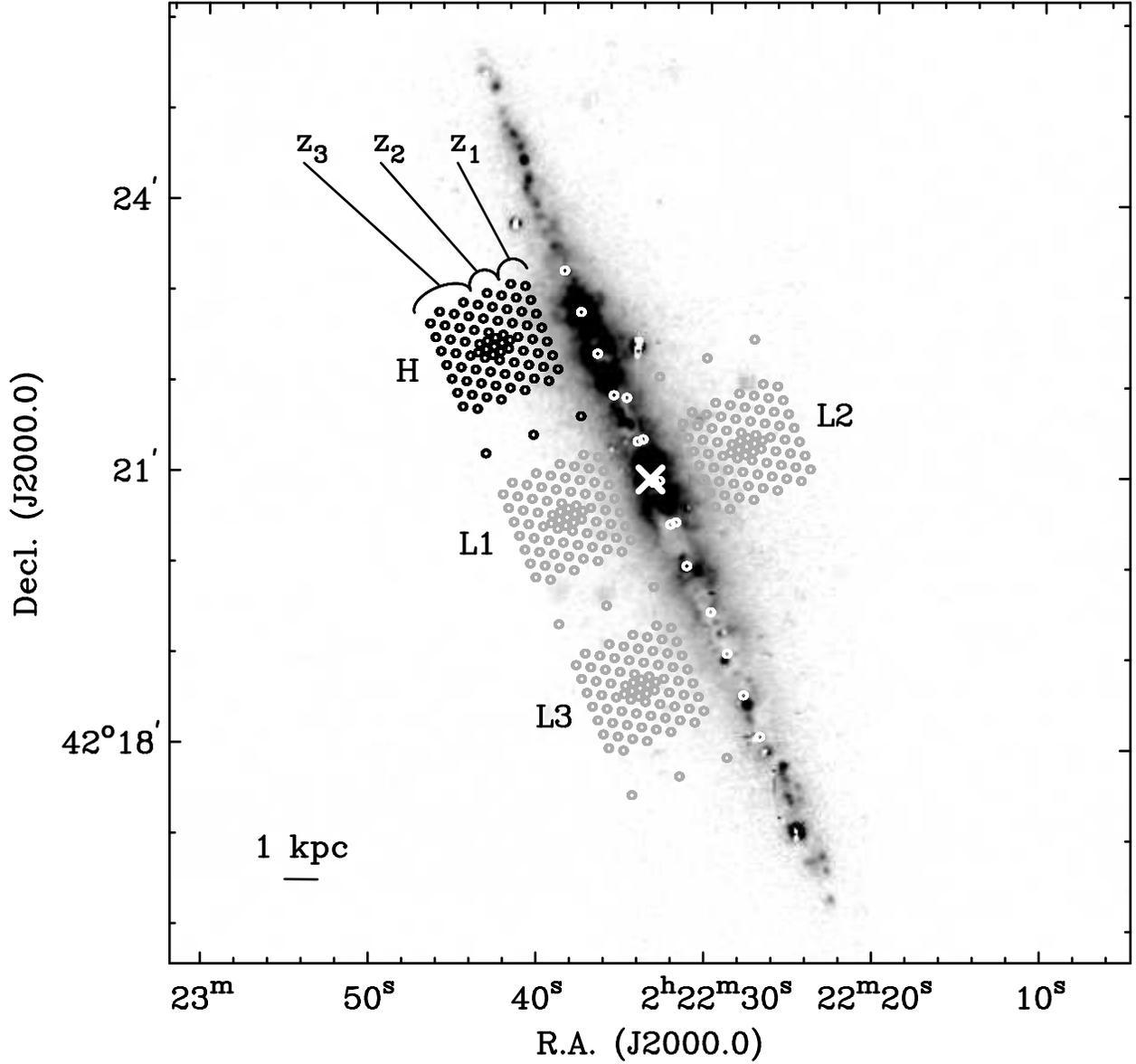}
\caption{The SparsePak pointings presented in this paper, overlaid on the \ha\ image of \citet{rkh90}. The spectrograph was set up in echelle mode for pointing H (\emph{black circles}) and in a lower spectral resolution mode for pointings L1, L2, and L3 (\emph{gray circles}). ``Sky'' fibers lying along the major axis are colored white for clarity. The rotational center of NGC 891 is marked with a white cross. The spatial scale (assuming $D=9.5\,\mathrm{Mpc}$) is shown in the lower left. Ranges of fibers used to construct PV diagrams (see text) are marked $z_1$ (\zi), $z_2$ (\zii), and $z_3$ (\ziii). These heights correspond to \zipc, \ziipc, and \ziiipc, respectively.}
\label{fig:pointings}
\end{figure}

Data were obtained during the nights of 2004 December 10--12 at the WIYN 3.5-m telescope. The SparsePak IFU \citep{bahrv04,bavwcs05} was used in conjunction with the Bench Spectrograph in two different configurations. For the first two nights, the echelle ($316\,\mathrm{lines\,mm^{-1}}$) grating was used at order 8, which provided a dispersion $0.205\,\mathrm{\AA\,pixel^{-1}}$ and a resolution $\sigma_{\mathrm{inst}} = 0.38\,\mathrm{\AA}$ ($\kms{17}$ at \ha); on the third night the $816\,\mathrm{lines\,mm^{-1}}$ grating was used at order 2, yielding a dispersion $0.456\,\mathrm{\AA\,pixel^{-1}}$ and a resolution $\sigma_{\mathrm{inst}} = 0.81\,\mathrm{\AA}$ ($\kms{37}$ at \ha). The grating angles for the two setups were 62.974$\degr$ and 51.114$\degr$, respectively. In both setups, the wavelength coverage included the \nii $\lambda\lambda\,6548,6583$, \ha, and \sii $\lambda\lambda\,6716,6731$ emission lines. The individual pointings of the fiber array are overlaid on an \ha\ image of NGC 891 \citep[from][]{rkh90} in Figure \ref{fig:pointings}. An observing log is presented in Table \ref{table:obslog}, and includes the ranges of radii ($R'$) and heights ($z$) covered by each pointing\footnote{To avoid confusion, we use $R'$ to indicate major axis distance, and $R$ to indicate galactocentric radius, throughout the paper.}.

The pointings shown in Figure \ref{fig:pointings} were selected based upon the following considerations. First, regions with prominent DIG emission are expected to be physically interesting since these areas host more active disk-halo flows. Moreover, bright EDIG is clearly preferable so that high signal-to-noise spectra may be obtained far from the plane. Because we are interested in the shapes of the velocity profiles, we need higher signal-to-noise than would be necessary to calculate velocity centroids. The analysis presented in \S\ \ref{subsection:modeling} requires spectra at large $R'$, where the rotation curve is flat (not rising), for the reasons described in that section. On the other hand, interesting kinematics may be observed along the minor axis (see \S\ \ref{subsection:dispersion}). Ideally, spectra would be obtained at the highest possible spectral resolution everywhere in the halo, but the necessary integration times make this prohibitive. Instead, we chose to observe at high spectral resolution where the DIG is brightest (the northeast quadrant), and at moderate spectral resolution in other regions of interest. The pointings were placed close enough to the plane to ensure DIG detection in all fibers, yet far enough to maximize our leverage on the determination of how rotation speeds vary with height.

It is important to note the details of the EDIG morphology observed at the locations of the SparsePak pointings (especially H and L3). As noted by both \citet{rkh90} and \citet{d90}, the EDIG emission is fainter in the southern half of NGC 891, possibly a consequence of a lower star formation rate in the underlying disk; pointing L3 is located in that region. In the northwest quadrant, \citet{rdwn04} detect thin extended filaments and loop structures atop the bright smooth component. In the northeast quadrant, the EDIG distribution appears to be largely smooth, with two prominent vertical filaments extending well away from the disk. We chose to place pointing H at the location of the latter filamentary structure. The perpendicular slit from \citet{r97} passes through the area covered by pointing H. With these choices of location for pointings H and L3, we observe regions of differing EDIG morphology.

Data reduction steps were performed in the usual way using the IRAF\footnote{IRAF is distributed by the National Optical Astronomy Observatories, which are operated by the Association of Universities for Research in Astronomy, Inc., under cooperative agreement with the National Science Foundation.} tasks CCDPROC and DOHYDRA. Cosmic-ray rejection was accomplished in the raw spectra with the package L.A.Cosmic \citep{v01}. The wavelength calibration was based on observations of a CuAr comparison lamp which was observed approximately once every $1.5\,\mathrm{hr}$ during the observing run. Spectrophotometric standard stars were observed throughout each night, and were used to perform the flux calibration. Because the standard stars were only observed with a single fiber, the flux calibration in other fibers is based on throughput corrections calculated from flat field exposures. The precision of our spectrophotometry was estimated by inspecting each flux calibrated standard star spectrum, which revealed variations at the 1\%\ level on the third night, and variations at the 10\%\ level on the first two nights. These variations result from slight errors in centering the standard stars on the central fiber, and from atmospheric seeing conditions scattering some of the starlight into adjacent fibers.

Subtraction of night-sky emission lines proved to be a difficult endeavor. The large angular size of NGC 891 prevented us from being able to use the dedicated sky fibers for their intended purpose, so an alternative procedure is needed. Because the response of the spectrometer to an unresolved emission line changes (slowly) as a function of fiber \citep{bavwcs05}, the night sky emission line spectrum cannot be satisfactorily subtracted from the object data by pointing the array at a patch of blank sky and averaging over all fibers; the changing line shape from fiber to fiber, intrinsic to the instrument, causes the average to be inappropriate for subtraction from the individual fibers. As indicated by \citet{bavwcs05}, one might adopt a ``beam-switching'' observing strategy, but at the cost of doubling the needed exposure times. A new, effective technique for subtracting sky lines based solely on the information in the object data has been developed \citep[see \S 6 of][]{bavwcs05}, but requires that the object emission be spread over a range of velocities. Since we observe only the approaching side of NGC 891 with pointing H, for example, there is not enough variation in the line velocities (as a function of fiber) to use this method. We were able, however, to use a modified version of this technique for pointing L3; see below.

We have attempted two alternative methods, not requiring object emission to be spread over a range of velocities, for removing sky lines which contaminate the line emission of interest. One method considers the shape of isolated sky lines located close to (and bracketing) the contaminating line, interpolates the observed line shape, and multiplies this interpolated sky line template by an appropriate factor (the factor begins near zero, and increases until negative fluxes result from the subtraction) to subtract the contaminating sky line. The other fits gaussian profiles to isolated sky lines and interpolates fit parameters to subtract the contaminating sky line (again, the amplitude, unknown because of blending with the line of interest, is increased until negative emission results). Both methods worked reasonably well, but left residuals behind. If we were simply interested in integrated line fluxes, the residuals would be acceptable. However, we seek information about the shape of the line, especially in the wings (see \S \ref{section:kinematics}), which may be significantly altered by the sky line. Therefore, the spectra presented here have been corrected for continuum emission, but the sky emission lines have in general not been subtracted. We note throughout the paper where sky lines are present in the spectra. In cases where we present sky-line subtracted spectra, the sky lines were originally located far enough out on the wings of the emission line of interest that a gaussian profile could be well fit and subtracted.

Final spectra were obtained by averaging the individual exposures at each pointing. The rms noise in each fiber was measured at each pointing; the mean and median of these values over all fibers are included in Table \ref{table:obslog} \citep[the rms noise in fibers toward the edges of the CCD is systematically higher than in the center of the chip, where the system throughput is higher; see][]{bavwcs05}. The final spectra were used to construct position-velocity (PV) diagrams at the three different heights indicated in Figure \ref{fig:pointings}, using the following procedure. The coordinates $R'$ and $z$ of each fiber were calculated in the galaxy frame [we take $R'$ positive on the receding (south) side]. The corresponding spectrum was then placed into the output PV diagram at the correct radius. We use radial bins of $4.7\arcsec$ (equal to the fiber diameter) so that the PV diagrams are largely continuous in $R'$. Note that at the largest $z$ (range $z_3$ in Fig. \ref{fig:pointings}), more than one fiber occupies each value of $R'$; overlapping spectra were averaged in this case, increasing the signal-to-noise.

Significant contamination by night sky emission lines has caused us to disregard the \ha\ line in pointings H and L3. In the former, the sky line falls on the side of the profile farthest from the systemic velocity, which is the part of the profile containing information about the rotation curve (see \S \ref{subsection:envelope}). In the latter, the sky line is at the same wavelength as the \ha\ line. In both cases, we instead consider the sum of the \nii $\lambda\,6583$ and \sii $\lambda\,6716$ lines. Note that in pointing L3, an additional sky line was located on the side of the \nii\ profile {\it closest} to systemic; this does not affect our ability to extract rotational information but should be kept in mind. To remove that line, we utilized the method developed by \citet{bavwcs05}. Because much of the object emission is at roughly the same wavelength, some oversubtraction occurred. However, we were able to utilize the farthest fiber from the disk, which received no object flux, to correct for this oversubtraction (which was the same in all fibers). Careful examination of the resulting subtracted spectra and the shapes of the subtracted sky lines confirmed that this approach works well for these observations. Any errors which may have crept into the final spectra will be mainly present on the side of the line profiles opposite from the critically important envelope side.

Our use of the forbidden lines \nii\ and \sii\ in this context deserves some comment, since the ratios \nii/\ha\ and \sii/\ha\ are well known in some DIG halos including NGC 891 to vary beyond what is expected based solely on photoionization. \citet{wm04}, for example, are able to reproduce most line ratios with photoionization alone, but additional heating sources (such as shocks) may still be required. Despite this possibility, the method used in \S\ \ref{subsection:modeling} in particular should be largely insensitive to such effects. The assumptions are that we can reproduce the density structure of the gas reasonably well, and that the kinematics of the gas are dominated by rotation. If these assumptions are reasonable, it is unlikely that localized energetic phenomena will significantly affect our results.

\section{Halo Kinematics}\label{section:kinematics}

In this section, we seek to extract information about the kinematics of the DIG in NGC 891 from the final spectra. The nearly edge-on viewing angle imposes the need to carefully consider the shape of the line profiles, which depend on both gas density and rotational velocity along the entire line of sight (LOS). We utilize two independent methods of extracting rotational information: in \S \ref{subsection:envelope}, we apply the envelope tracing method \citep[e.g.,][]{sr01,sa79} to the observed line profiles; in \S \ref{subsection:modeling}, we develop a three-dimensional model of the galaxy for comparison with the data. Data from pointings H and L3 are considered in this section, because the radii included are large enough that the rotation curve is approximately flat. Pointings L1 and L2, which lie along the minor axis, contain emission from gas at radii where the rotation curve is still rising. As pointed out by \citet{foss05}, at these small radii changes in the rotation curve and the density profile cannot be distinguished. Therefore, we cannot robustly extract rotational information from those spectra. We come back to those pointings later in the paper.

\subsection{Envelope Tracing Method}\label{subsection:envelope}

The envelope tracing method works under the assumption of circular rotation. A LOS crossing through such a disk will intercept gas at many different galactocentric radii, but at the line of nodes (the line in the plane of the sky where $R=R'$), the projection of the velocity vector on the LOS is maximized. Thus, the highest velocity (relative to the systemic velocity, and after a correction for velocity dispersion has been made) in a velocity profile at $R'$ is taken to be the value of the rotation curve at galactocentric radius $R=R'$. By following this procedure for every $R'$ in a PV diagram, a rotation curve $V_{\mbox{\scriptsize{rot}}}(R)$ is built up. The details of our algorithm for implementing this method are described in Paper I and are not repeated here. We note, however, the values used for the envelope tracing parameters (refer to eqns. 1 and 2 in Paper I): $\eta = 0.2$, $I_{\mathrm{lc}}=3\sigma$ (where $\sigma$ refers to the rms noise), and $(\sigma_{\mathrm{inst}}^2+\sigma_{\mathrm{gas}}^2)^{1/2}=\kms{20}$. The latter quantity, determined empirically during the modeling process described in \S\ \ref{subsection:modeling}, adds a constant offset to all of the derived rotation speeds; as long as $\sigma_{\mathrm{gas}}$ is approximately constant with $z$, our determination of the \emph{change} in rotation curve with height is insensitive to the adopted value. In any case, the instrumental broadening dominates for these observations.

\begin{figure}
\plotone{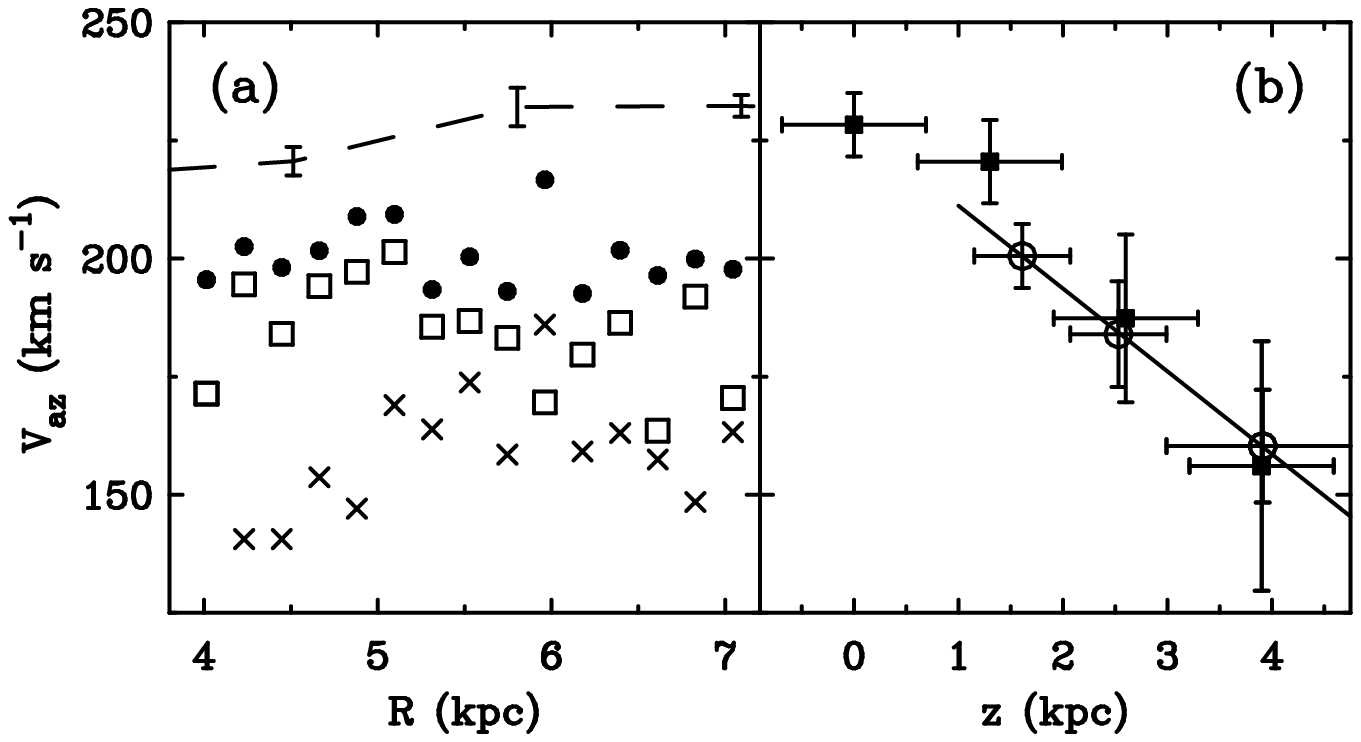}
\caption{Results of applying the envelope tracing method to PV diagrams constructed from the spectra obtained in pointing H (northeast quadrant). (a) Rotational velocities are shown for \zipc\ (\emph{solid circles}), \ziipc\ (\emph{empty squares}), and \ziiipc\ (\emph{crosses}). A major axis rotation curve determined from \hi\ observations is included for reference (\emph{dashed line}). (b) The average rotational velocity at each height is shown for the DIG (\emph{empty circles}) and for the \hi\ (for radii included in pointing H; \emph{solid squares}). Horizontal errors for the DIG data indicate the range of $z$ covered by the fibers used to construct the individual PV diagrams; those for the \hi\ data reflect the angular resolution ($30\arcsec = 1.4\,\mathrm{kpc}$). The best-fit solution for the DIG ($dv/dz=17.5\pm \kmskpc{5.9}$), described in the text, is plotted for reference (\emph{solid line}). The \hi\ data were kindly provided by F. Fraternali.}
\label{fig:env_h}
\end{figure}

The envelope tracing method was applied to the PV diagrams extracted from pointing H, at each of the three $z$-heights. The results are shown in Figure \ref{fig:env_h}a. It seems clear from visual inspection that a vertical gradient in rotational velocity is present. To derive the value of this gradient, two methods were used. First, at each height, the mean rotational velocity was determined (see Figure \ref{fig:env_h}b). A linear fit to those points revealed a gradient with value $dv/dz=17.5\,\pm\,\kmskpc{5.9}$ (throughout this paper, to be consistent with previous studies, we use the symbol $dv/dz$ to mean the {\it magnitude} of the gradient as in eq. \ref{eqn:dvdz}; it actually represents a negative quantity because the halo is rotating slower than the disk). The best-fit line is shown in Figure \ref{fig:env_h}b. We also calculated a linear fit to the velocities (as a function of height) at each $R$, and took the average of the gradients determined at each radius; that value was $dv/dz=17.6\,\pm\,\kmskpc{5.3}$. The large error bar reflects the fact that the gradient appears to be somewhat steeper at lower $R$.

In principle, we could use the linear fits described above, in conjunction with information about the major axis rotation curve, to constrain the starting height of the gradient, $z_{\mathrm{cyl}}$. However, our choices of the envelope tracing parameters (in particular, $I_{\mathrm{lc}}$) are somewhat arbitrary, and lead to an uncertainty in the zero-point scaling of the velocities derived from the method. This issue does not affect our ability to determine the value of $dv/dz$, but an error in the zero-point offset would translate directly to an error in $z_{\mathrm{cyl}}$. In \S\ \ref{subsection:modeling}, we utilize a more powerful method of examining the halo kinematics, which also leads to more robust constraints on the value of $z_{\mathrm{cyl}}$ than can be provided by the envelope tracing method.

\subsection{PV Diagram Modeling}\label{subsection:modeling}

In the envelope tracing method, the rotation speed determined within each velocity profile of a PV diagram lies between the intensity peak of the profile, and the part of the profile at $3\sigma$ \citep[see][]{sr01}. One could imagine using the envelope tracing method on the same velocity profiles but with increased noise values. As the noise value increases relative to the peak of the profile, the part of the profile at $3\sigma$ moves closer to the profile peak, and the location of the rotation speed will thus move away from the true rotational velocity and toward the systemic velocity. Therefore, since the signal-to-noise ratio decreases with increasing $z$, what we take to be a rotation velocity gradient in Figure \ref{fig:env_h} could be an artifact of how the envelope tracing method determines rotation velocities. To check that our result is robust, we have generated galaxy models in order to analyze the halo kinematics with a method which takes the noise into account. We begin by generating models intended to match the data from pointing H only, and later attempt to model the data from pointing L3 separately.

The galaxy models described in this section are generated with a modified version of the Groningen Image Processing System \citep[GIPSY;][]{vtbzr92} task GALMOD. The modifications include the addition of a vertical gradient in rotation velocity, resulting in rotation speeds of the form
\begin{equation}
\label{eqn:dvdz}
V_{\mathrm{rot}}(R,z)=
\left\{
\begin{array}{ll}
V_{\mathrm{rot}}(R,z=0) & \mbox{for $z\leq z_{\mathrm{cyl}}$} \\
V_{\mathrm{rot}}(R,z=0)-\frac{dv}{dz}[z-z_{\mathrm{cyl}}] & \mbox{for $z>z_{\mathrm{cyl}}$}
\end{array}
\right.
\end{equation}
where $V_{\mathrm{rot}}(R,z=0)$ is the major axis rotation curve, $dv/dz$ is the magnitude of the vertical gradient in rotation velocity, and $z_{\mathrm{cyl}}$ is the height at which the gradient begins (below that height, the halo rotates cylindrically). To keep the number of free parameters to a minimum, given our limited information, we assume that $dv/dz$ is constant with $z$. This assumption appears to be justified based on the results of the envelope tracing method (see Figure \ref{fig:env_h}b).

\begin{figure}
\plotone{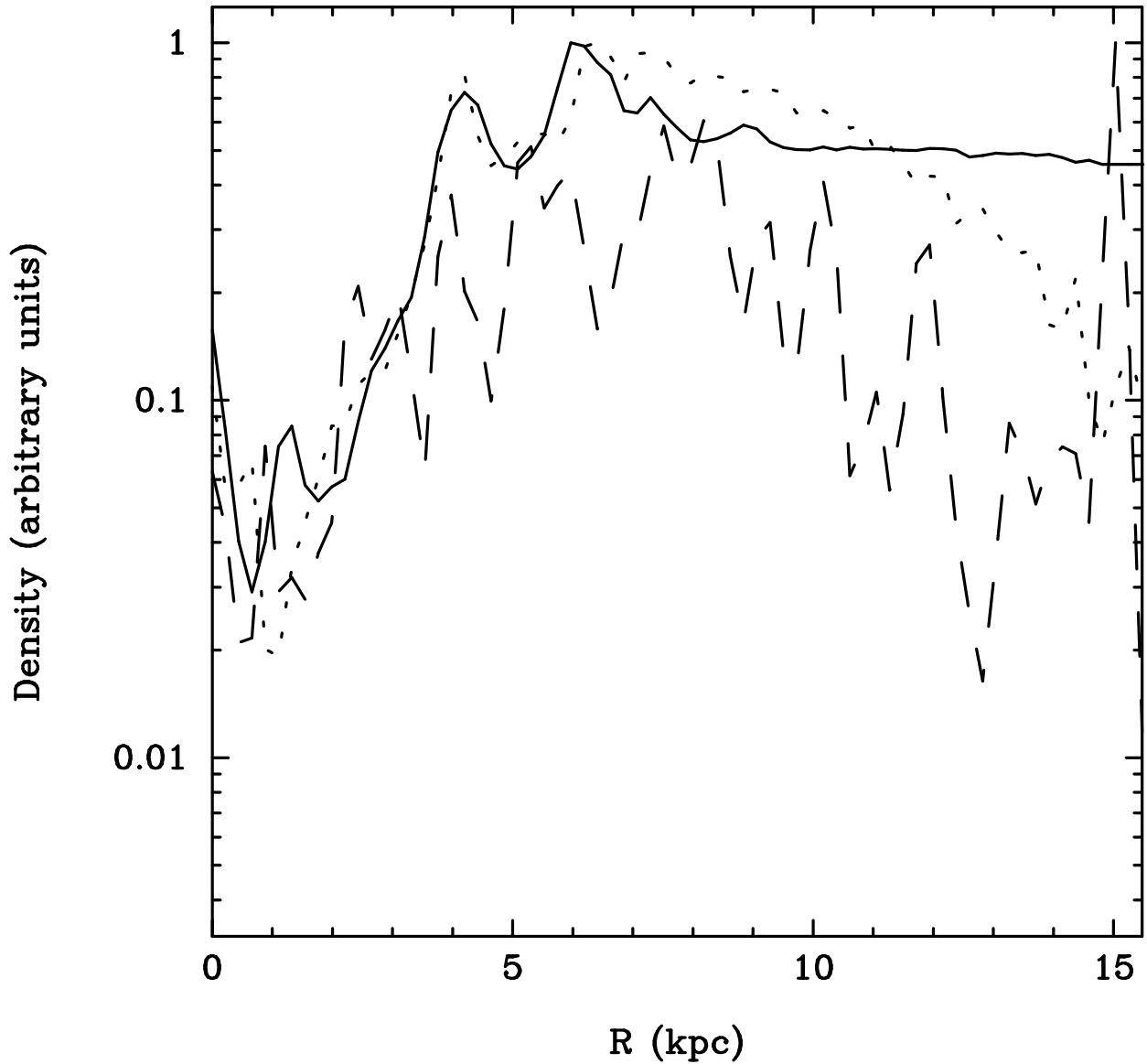}
\caption{Profiles of gas density as a function of galactocentric radius which are used to generate the models described in the text. Profiles are shown for \zipc\ (\emph{solid line}), \ziipc\ (\emph{long-dashed line}), and \ziiipc\ (\emph{dotted line}). Because we seek to match the signal-to-noise ratio in the model to that in the data, the actual density values are of little importance. Each profile has been normalized to its peak for presentation.}
\label{fig:densprofs}
\end{figure}

The model inputs include the distance, inclination, and systemic velocity of NGC 891 (these are set to the values discussed in \S\ \ref{section:introduction}), as well as the velocity dispersion, radial density profile, and major axis rotation curve. The best value of the velocity dispersion was found to be $\kms{20}$ (note that this value includes the instrumental broadening as well as the gas dispersion, and is strictly appropriate only for pointing H); this determination of the velocity dispersion motivated our use of the same value in the envelope tracing method (see \S\ \ref{subsection:envelope}). To estimate the radial density profile, cuts were made through the \ha\ image of \citet{rkh90}, along slices parallel to the major axis and at heights corresponding to the heights of the PV diagrams constructed from the SparsePak data. The GIPSY task RADIAL was used to produce radial density profiles that would result in these intensity cuts. The amplitude of the model radial density profiles is set so that the model signal-to-noise ratio (the artificial observations created by GALMOD include noise) is matched as well as possible to that in the data. A flat rotation curve was used initially, and the radial density profile was adjusted until the appearance of the PV diagrams was well matched. A flat rotation curve was sufficient to obtain an excellent match to the data, and was therefore used for the remaining modeling (we select $V_{\mathrm{rot}}(z=0)=\kms{230}$, but this value cannot be uniquely constrained for this method with our data; see below). The best fit radial density profiles are shown in Figure \ref{fig:densprofs}. By specifying a different radial density profile for each height, we hope to match the true gas distribution at each location with greater success than by assuming that the shape of the radial density profile is constant with $z$.

The output models take the form of RA-Decl.-$v_{\mathrm{hel}}$ data cubes. We have written a GIPSY script which performs artificial SparsePak observations of the output data cubes. Specifically, it extracts spectra along sight lines, arranged in the pattern of the SparsePak fibers on the sky, through the data cube. Thereafter, PV diagrams are created from the actual SparsePak data and the modeled SparsePak data in the same way. We proceed by generating a grid of models, constructing difference PV diagrams, and selecting the models that minimize the absolute mean difference and rms difference. Using this method, we are able to specify a best-fit value of $dv/dz$, but without a direct measurement of the major axis rotation curve, we cannot uniquely specify $z_{\mathrm{cyl}}$. Instead, we obtain a relationship between the rotation speed at $z=0$ and the starting height of the gradient.

Using the procedure described above, an initial exploration of parameter space suggested $dv/dz=\kmskpc{14.9}$. To constrain the value of $z_{\mathrm{cyl}}$, we consider the range of major axis rotation speeds allowed by the \hi\ observations \citep{foss05}, roughly 220 to $\kms{235}$. Those values correspond to $z_{\mathrm{cyl}}=1.6$ and $0.6\,\mathrm{kpc}$, respectively. For our adopted value $V_{\mathrm{rot}}(z=0)=\kms{230}$, we obtain $z_{\mathrm{cyl}}=0.9\,\mathrm{kpc}$. To ensure consistency, another grid of models, differing only in the value of $dv/dz$, was generated using the adopted values of $V_{\mathrm{rot}}(z=0)$ and $z_{\mathrm{cyl}}$; the best values of $dv/dz$ were $\kmskpc{15.2}$ and $\kmskpc{15.4}$ for \ziipc\ and \ziiipc, respectively.

\begin{figure*}
\epsscale{.75}
\plotone{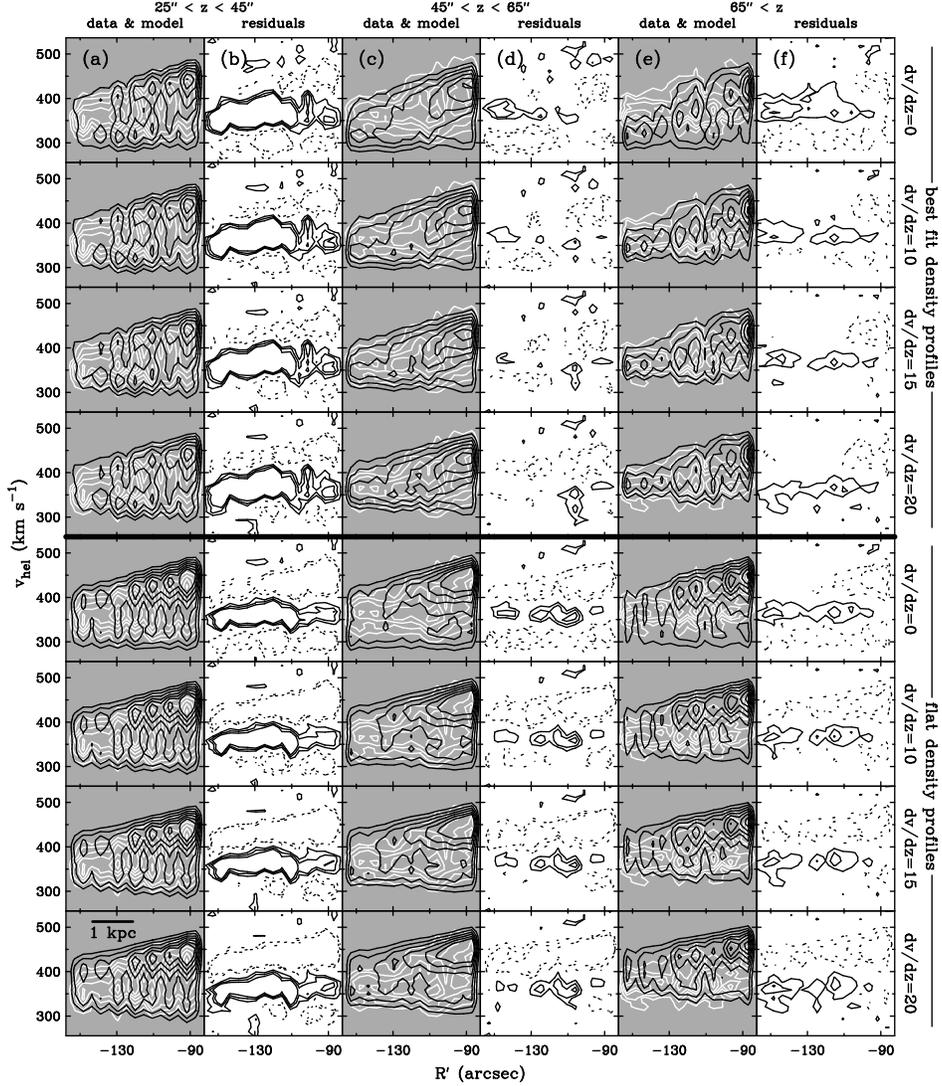}
\epsscale{1}
\caption{Comparison between PV diagrams constructed from the SparsePak data and from the galaxy models described in \S \ref{subsection:modeling}. In the left panels (a, c, and e) of each pair of columns, the data are shown with white contours, and the models are displayed with black contours. In the right panels (b, d, and f), the difference between data and model is shown. The leftmost columns (a and b) are for heights \zi; the central columns (c and d) for \zii; and the rightmost columns (e and f) for \ziii. The top four rows include models constructed using the best-fit radial density profiles shown in Figure \ref{fig:densprofs}; the bottom four rows include models constructed using flat radial density profiles. The azimuthal velocity gradient used in the models in each row is listed on the right edge. From left to right, the contour levels in each column are (a) 10$\sigma$ to 40$\sigma$ in increments of 5$\sigma$; (b) 3, 5, and 7$\sigma$ (positive for solid contours, negative for dashed contours); (c) 5$\sigma$ to 20$\sigma$ in increments of 3$\sigma$; (d) the same as in (b); (e) 3$\sigma$ to 12$\sigma$ in increments of 1.5$\sigma$; and (f) 2, 4, and 6$\sigma$ (positive for solid contours, negative for dashed contours). The systemic velocity is $\kms{528}$. $|R'|$ increases to the north. The angular size corresponding to $1\,\mathrm{kpc}$ is shown in the bottom left panel.}
\label{fig:biggrid}
\end{figure*}

The models described above are completely azimuthally symmetric. In reality, NGC 891 is certainly not azimuthally symmetric, due to the presence of spiral arms, filamentary halo structures, \hii\ regions, and so on. To check that our results are not biased by a poor specification of the radial density profile, we have also constructed a grid of models with a completely flat radial density profile (see the lower four rows of Figure \ref{fig:biggrid}). Although this choice of density profile poorly reproduces the shape of the data PV diagrams, the best match occurs once again for the model with $dv/dz\approx \kmskpc{15}$. This suggests that errors in our best-fit radial density profiles will not significantly alter the results that we have derived for the halo kinematics. Put another way, the changes in the PV diagrams as a function of $z$ are dominated by the decrease in rotational velocity with height.

Although we have presented strong evidence for a vertical gradient in azimuthal velocity with a magnitude very close to $\kmskpc{15}$, this result is only valid for the region of the halo of NGC 891 covered by pointing H \citep[in the northeast quadrant -- incidentally, this falls within the same region examined by][]{foss05}. Can this result be extended to describe the kinematics elsewhere in the halo? Unfortunately, we have not obtained high spectral resolution SparsePak data covering the entire halo of this galaxy, but a lower spectral resolution pointing (L3) has been obtained in the southeast quadrant. We now attempt to model that quadrant of NGC 891 to test whether the kinematics are consistent with the results from pointing H.

\begin{figure*}
\plotone{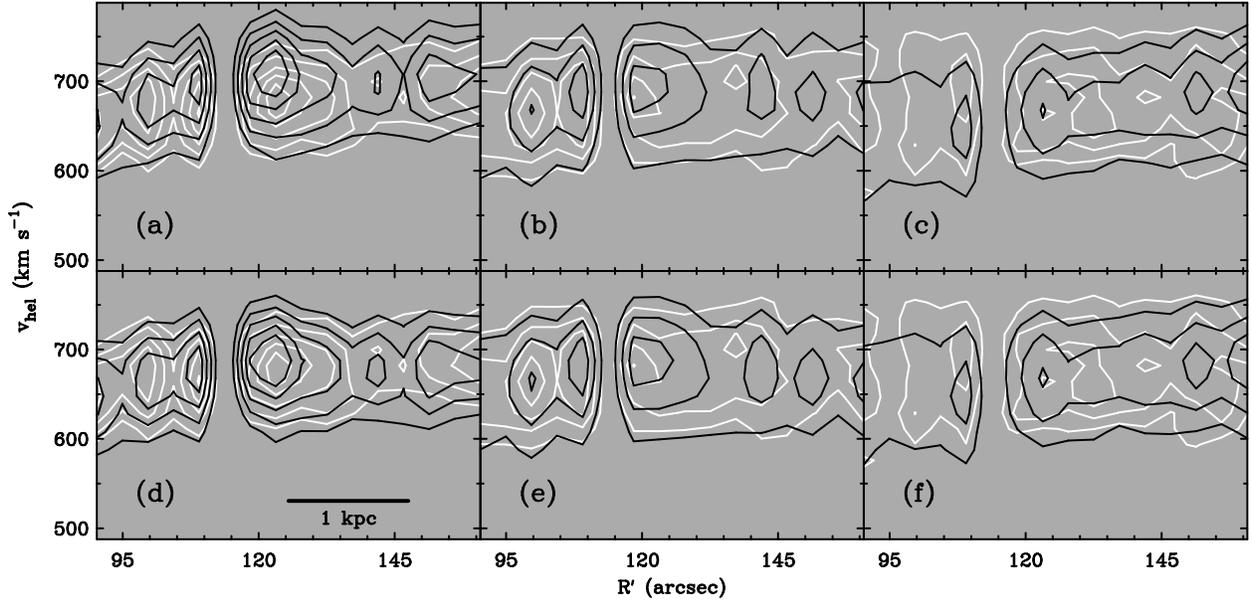}
\caption{Comparison between PV diagrams constructed from pointing L3 (\emph{white contours}), and two models: one including a gradient $\kmskpc{15}$ as described in the text (\emph{black contours}, top panels), and the other including a constant rotation speed with height $V_{\mathrm{rot}}(R,z)=\kms{175}$ (\emph{black contours}, bottom panels). Contour levels are (a,d) 10$\sigma$ to 30$\sigma$ in increments of 5$\sigma$ for \zi; (b,e) 5$\sigma$ to 20$\sigma$ in increments of 5$\sigma$ for \zii; and (c,f) 3$\sigma$ to 9$\sigma$ in increments of 3$\sigma$ for \ziii. Pointing L3 is on the receding side of the galaxy; the systemic velocity is $\kms{528}$. $R'$ increases to the south. The angular size corresponding to $1\,\mathrm{kpc}$ is shown in the bottom left panel.}
\label{fig:l3compare}
\end{figure*}

A galaxy model appropriate for the region covered by pointing L3 was produced in the manner described above. Initially, the best-fit model parameters obtained for pointing H were considered, with the exception that the dispersion velocity was corrected for the instrumental broadening, and set to $\kms{40}$. In Figure \ref{fig:l3compare}, we display a comparison between the data and a model with a constant gradient $dv/dz=\kmskpc{15}$ (top panels). Such a model is clearly inadequate to match the data. Inspection of the PV diagrams shows that the rotation speeds at each height are not consistent with halo kinematics of the form in equation \ref{eqn:dvdz}. Rather, the data indicate a halo with a negligible velocity gradient, but with a constant rotation velocity ($\kms{\sim 170-180}$) much slower than that of the underlying disk. In the bottom panels of Figure \ref{fig:l3compare}, we compare the data to a model with a constant rotation speed $V_{\mathrm{rot}}(R,z)=\kms{175}$, and no gradient at all. This second model clearly gives much better agreement. We conclude that the kinematics in the southeast quadrant of NGC 891 are markedly different than those in the northeast quadrant, and are well described by a constant velocity of approximately $\kms{175}$.

\section{The Ballistic Model}\label{section:ballistic}

The physical cause for the vertical gradient in rotation velocity measured in NGC 891 is not well understood. To be sure, two independent models have been shown by other researchers to predict such a gradient and reproduce its magnitude: a hydrostatic model \citep{bcfs06} and a hydrodynamic model of gas accretion during disk formation \citep{kmwsm05}. Yet neither model allows for the apparent connection between star formation activity in the disk and the prominence of gaseous halos \citep[e.g.,][]{mv03,rd03}. The evidence seems to favor a physical situation similar to that described by the fountain model \citep{sf76,b80}. As a first step toward realizing such a model, we have developed the ``ballistic model'' \citep[Paper I;][]{cbr02}. We note that \citet{fb06} have developed a similar model to ours, and have recovered similar results.

The ballistic model is described elsewhere; only a brief description will be given here. The model numerically integrates the orbits of clouds in the galactic potential of \citet{wmht95}. The clouds are initially located in an exponential disk with a scale length $R_0$ set to match the observations, and are launched into the halo with an initial vertical velocity randomly selected to be between zero and a maximum ``kick velocity'' $V_\mathrm{k}$, which effectively sets the vertical scale height of the halo density distribution, and can thus be constrained by observations. As the clouds move upward out of the disk, they feel a weaker gravitational potential and migrate radially outward; in order to conserve angular momentum, the rotational velocity of the clouds decreases. The ballistic model therefore naturally produces a vertical gradient in azimuthal velocity. The specific parameters that form the ballistic ``base model'' for NGC 891 \citep[as determined by][]{cbr02} are $R_0=7\,\mathrm{kpc}$, $V_\mathrm{k}=\kms{100}$, and the circular velocity $V_\mathrm{c}=\kms{230}$.

\subsection{Rotation velocity gradient}\label{subsection:gradient}

\begin{figure}
\epsscale{0.5}
\plotone{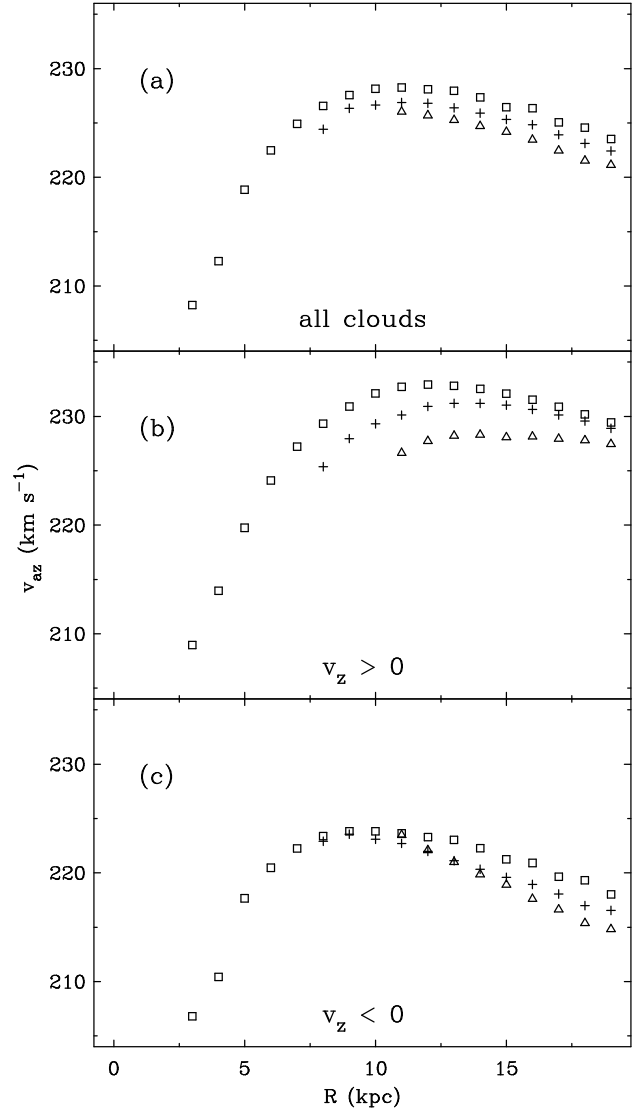}
\caption{Azimuthal velocities extracted from the ballistic base model, for (a) all clouds, (b) clouds moving upward (away from the disk), and (c) clouds moving downward (toward the disk). The rotation curves are averages computed within \zipc\ (\emph{squares}), \ziipc\ (\emph{plusses}), and \ziiipc\ (\emph{triangles}).}
\label{fig:vazb}
\end{figure}

The individual orbits of clouds in the ballistic model can be directly examined to explore the predicted variation in rotational velocity with height. We have directly extracted the average rotation curve at three heights, corresponding to the ranges of $z$ considered in \S\ \ref{section:kinematics}: \zipc; \ziipc; and \ziiipc. These rotation curves are shown in Figure \ref{fig:vazb}a. Note that the lack of measured rotational velocities at inner radii is because no clouds at these radii reach these heights in the model. The evacuated region in the model is roughly cone-shaped; therefore the range of radii with no measured velocities increases with height. We come back to the issue of radial redistribution of clouds in \S\ \ref{subsection:migration}.

Figure \ref{fig:vazb}a demonstrates that a vertical gradient in azimuthal velocity is indeed present in the ballistic model, but the magnitude of the gradient is extremely shallow. The gradient, averaged over all radii where data are present at all three heights, is $\kmskpc{1.1\pm 0.1}$. In comparison to the gradient measured in pointing H ($\kmskpc{15-18}$), the ballistic model prediction is too shallow by over an order of magnitude.

We also consider the effect of counting only clouds which are moving upwards in the ballistic model, and only clouds which are moving downwards. Physically, these cases would correspond to a fountain flow in which only the gas leaving (returning to) the plane is ionized. In the latter case (see Figure \ref{fig:vazb}c), the average gradient is even shallower: $\kmskpc{1.0\pm 0.5}$. Even in the upward-moving case, the gradient is only $\kmskpc{1.7\pm 0.6}$. Clearly, the ballistic model in its present form is inadequate to explain the observed kinematics in the halo of NGC 891 at pointing H. There is little or no gradient above $z=1.2\,\mathrm{kpc}$ at pointing L3, but the large decrease in azimuthal velocity that we conclude occurs between $z=0$ and $z=1.2\,\mathrm{kpc}$ is definitely inconsistent with the ballistic model.

\subsection{Emission profiles}\label{subsection:migration}

The ballistic model also makes strong predictions regarding the steady-state spatial distribution of clouds in the halo. In the simulation, a cloud with a fixed initial vertical velocity will reach higher $z$ when initially placed at larger $R$, where the gravitational potential is weaker (see, for example, Fig. 14 in Paper I). Because clouds in the outer disk orbit higher than clouds in the inner disk, the halo is more vertically extended at larger $R$. Note that this behavior manifests itself regardless of the level of radial migration which may be predicted by the model. When viewed from an edge-on perspective, a halo with such a density structure would show distinctive emission profiles parallel to the major axis. Are such profiles actually observed?

\begin{figure}
\epsscale{0.4}
\plotone{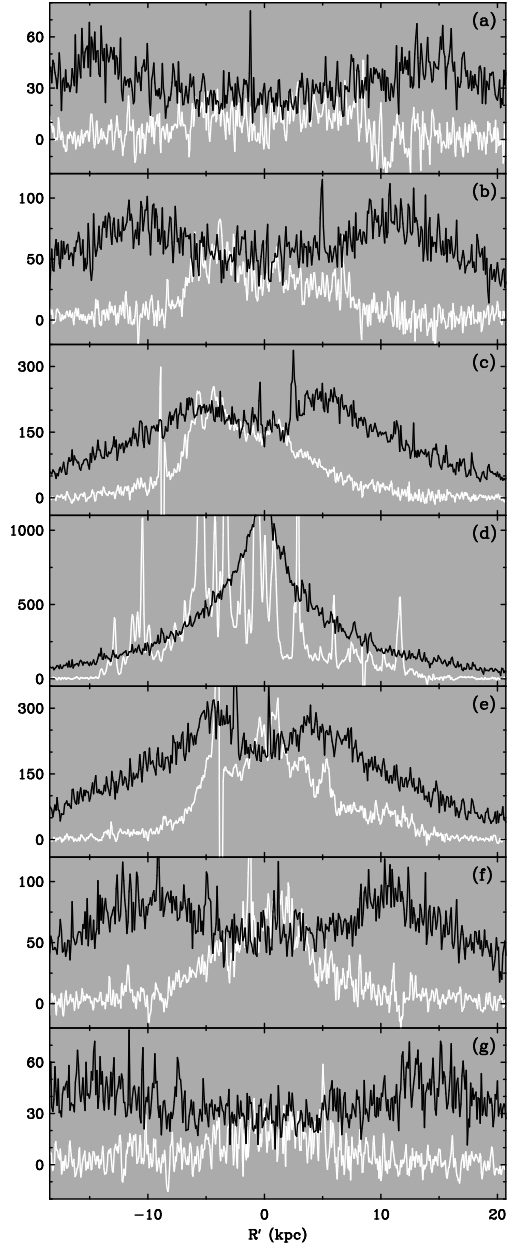}
\caption{Intensity profiles along cuts through the ballistic model (\emph{black lines}) and the \ha\ data (\emph{white lines}). The vertical scale is in units of EM ($\mathrm{pc\,cm^{-6}}$) for the \ha\ data and number of clouds (multiplied by a constant for presentation) for the ballistic model data. The cuts were made parallel to the major axis at heights (a) $z=3\,\mathrm{kpc}$ (to the east of the disk); (b) $z=2\,\mathrm{kpc}$ (east); (c) $z=1\,\mathrm{kpc}$ (east); (d) $z=0\,\mathrm{kpc}$; (e) $z=1\,\mathrm{kpc}$ (west); (f) $z=2\,\mathrm{kpc}$ (west); and (g) $z=3\,\mathrm{kpc}$ (west). The sky coordinate $R'$ increases to the south.}
\label{fig:intcuts}
\end{figure}

Because we are only interested in total \ha\ intensity as a function of radius, and to increase the area over which we can make a comparison, we consider radial cuts through the \ha\ image from \citet{rkh90}. These intensity cuts are compared to similar cuts made through a total intensity map generated from the output of the ballistic model, and are shown in Figure \ref{fig:intcuts}. We note that we are only interested in the shape of the intensity cuts, so the absolute values are unimportant. We also note that in the model, each cloud is assumed to provide equal intensity; therefore, the expected total intensity scales linearly with the column density of clouds.

Inspection of Figure \ref{fig:intcuts} shows a lack of agreement between the data and the model. Cuts through the ballistic model show a strong central depression which grows with height. For example, the emission at $z=\pm\,1\,\mathrm{kpc}$ peaks in the radial range $R'\lesssim 5\,\mathrm{kpc}$, but at $z=\pm\,3\,\mathrm{kpc}$, the peaks are located at $R'\approx 15\,\mathrm{kpc}$. In contrast, the \ha\ profiles show a general decline in intensity with increasing radius at all heights. An exception is perhaps seen on the east side of the disk, though the central depression is not nearly as pronounced as in the model. The data do not seem to indicate a significant change in the shape of the radial density profile with height. The behavior of the ballistic model is caused by the weakening of the gravitational potential with increasing radius; it is therefore unlikely that any realistic mass model would lead to different results. This large discrepancy indicates a further failure of the ballistic model. The relatively small variation of the profiles in Figure \ref{fig:intcuts} with $z$ suggests a hydrodynamic effect which regulates the vertical flow.

\subsection{Minor-axis velocity dispersion}\label{subsection:dispersion}

\begin{figure}
\plotone{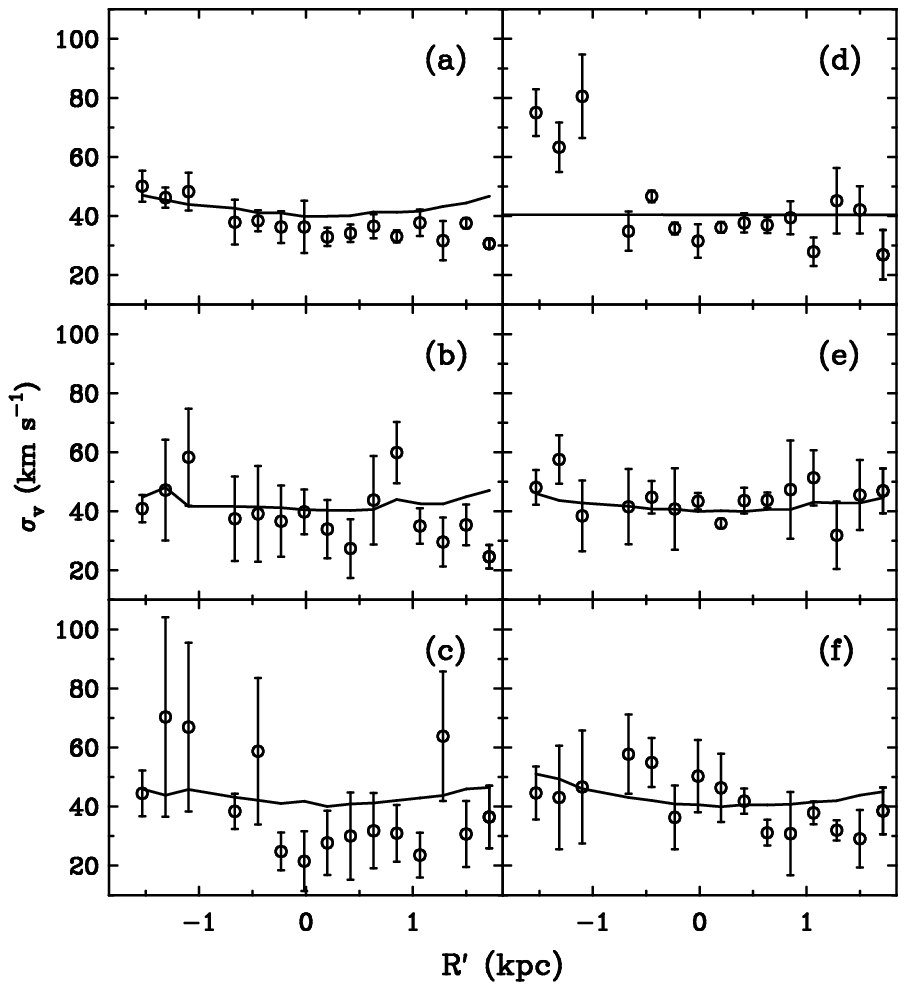}
\caption{\ha\ velocity dispersions (\emph{open circles}) from pointings L1 (a--c) and L2 (d--f), and velocity dispersions measured from the best fit model described in \S\ \ref{subsection:modeling} (\emph{solid lines}), which was modified to correct for the instrumental broadening (making the dispersion in the model $\kms{40}$). Widths were calculated using PV diagrams constructed at heights (a,d) \zipc; (b,e) \ziipc; and (c,f) \ziiipc. Errors on the model line widths are typically about $\kms{0.5-1}$. For $R' > +0.9\,\mathrm{kpc}$, the \ha\ widths should be considered lower limits due to confusion with a sky line. The exceptionally high data points in (d) are due to confusion with another, fainter sky line, and should be disregarded.}
\label{figure:widths}
\end{figure}

Along the minor axis, the rotational velocity vectors should be perpendicular to the LOS (assuming circular rotation). On the other hand, any radial motions would be parallel to the LOS, and would thus contribute to the width of the velocity profiles. We therefore examine the velocity dispersions in pointings L1 and L2 to search for evidence of radial motions along the LOS. First, we make artificial observations of the best-fit model from \S\ \ref{subsection:modeling} (generated with velocity dispersion $\kms{40}$ to account for the lower spectral resolution of these pointings). Velocity widths were calculated in the same way for the \ha\ line in these pointings and from artificial observations of the model. Note that in making this comparison, we assume the axisymmetry of the best-fit model. The results are shown in Figure \ref{figure:widths}.

It appears that the observed velocity dispersions in the data are roughly consistent with the velocity dispersion in the model. It should be stressed that we have determined the value of the gas dispersion from pointing H, and have only adjusted the model inputs to account for a different value of instrumental broadening. At the lowest heights (\zi), the modeled velocity dispersions are slightly higher than those in the data; it seems unlikely that there are any significant radial motions at these locations in the halo. Elsewhere, the large uncertainties in the data could allow for additional radial motions at the level of $v_{\mathrm{rad}}\leq \kms{30}$.

\begin{figure}
\plotone{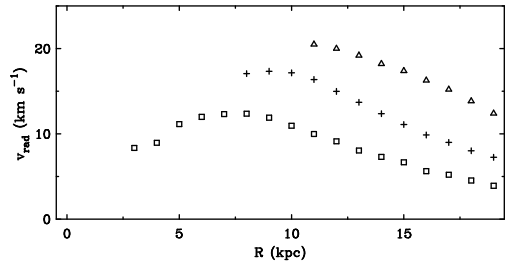}
\caption{Average radial velocities, as a function of galactocentric radius, predicted by the ballistic base model. Radial velocities were computed within height ranges \zipc\ (\emph{squares}); \ziipc\ (\emph{plusses}); and \ziiipc\ (\emph{triangles}). Missing data points at low radius correspond to a lack of clouds at those locations.}
\label{figure:bmvrad}
\end{figure}

What level of radial motion is predicted by the ballistic model? In Figure \ref{figure:bmvrad}, we display the average radial motions of individual clouds as a function of radius in the three height ranges considered here. In the model, radial velocities are higher farther from the disk, and range from approximately $\kms{5}$ to $\kms{20}$ at the largest heights. The data presented in Figure \ref{figure:widths} may just be consistent with radial motions of this magnitude. The large uncertainties in the velocity dispersions measured from the data do not allow us to place stronger constraints on possible radial motions. Observations of the same region with higher spectral resolution (as in pointing H) would allow us to better understand how radial motions contribute to the minor axis velocity dispersions.

Note that even hydrostatic models, like those of \citet{bcfs06}, in general will require an increasing velocity dispersion as a function of height to preserve hydromagnetic stability. Hydrostatic models of \citet{bc90}, for example, predict a velocity dispersion increasing from $\kms{30}$ at $1\,\mathrm{kpc}$ to $\kms{60}$ at $4\,\mathrm{kpc}$. Our data also provide constraints on this class of models.

\subsection{Halo potential}\label{subsection:phihalo}

In previous papers, it was shown that the exact form of the dark matter halo potential is of little importance in shaping the orbits of the clouds in the ballistic model. Here, as an aside, we examine the relative importance of contributions to the total galactic potential from the disk, bulge, and halo. The vertical and radial components of the gravitational acceleration are calculated numerically on a grid of $R$ and $z$, using $g_{R,i}(R,z)=-\partial \phi_{i}(R,z)/\partial R$ and $g_{z,i}(R,z)=-\partial \phi_{i}(R,z)/\partial z$, where the index $i$ indicates that the disk, halo, and bulge contributions are considered separately. Finally, the total barycentric gravitational acceleration is calculated: $g_i(R,z)=(g_{R,i}^2+g_{z,i}^2)^{1/2}$.

\begin{figure}
\epsscale{0.5}
\plotone{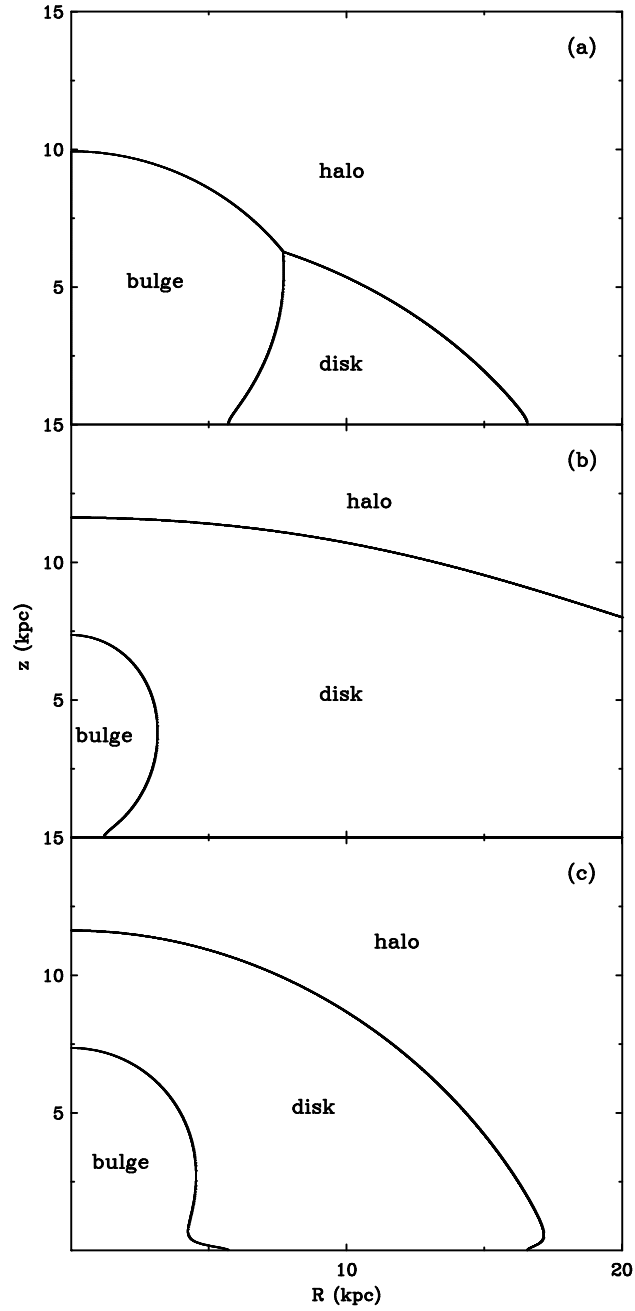}
\caption{Depictions of the regions in the ballistic model within which the disk, bulge, and halo potentials contribute the most to the (a) radial component of the gravitational acceleration; (b) the vertical component of the gravitational acceleration; and (c) the total gravitational acceleration.}
\label{fig:compmaps}
\end{figure}

Figure \ref{fig:compmaps} shows the regions where the disk, bulge, and halo potentials provide the maximum contribution to the radial, vertical, and total gravitational acceleration. Clearly, the halo potential is crucial in driving the dynamics at large radii (which is why the halo potential is incorporated in the first place) and large heights, but the disk and/or bulge potentials remain dominant within the $R$ and $z$ through which most clouds travel \citep[for orbits with initial radii $R=4$, 8, 12, and $16\,\mathrm{kpc}$, the maximum heights reached are approximately $z\approx 1.2$, 2.5, 4.5, and $6.5\,\mathrm{kpc}$ respectively; see Figure 3 of][]{cbr02}. At the largest radii and heights, the halo begins to become the dominant factor, but in general the disk and bulge potentials are most important. Thus the rotational velocities which we extract from the ballistic model are rather insensitive to the shape (spherical or flattened) of the dark matter halo. We note that the same insensitivity to halo shape is also observed in the model of \citet{fb06}, who use a different formulation of the galactic potential.

\section{Conclusions}\label{section:conclusions}

We have presented SparsePak observations of the gaseous halo of the edge-on NGC 891. Spectra from the individual fibers that make up the SparsePak array were arranged into PV diagrams, which were then analyzed in two separate ways to investigate the rotation field of the halo gas. First, rotational speeds were directly extracted from the PV diagrams using the envelope tracing method, revealing a vertical gradient in azimuthal velocity with magnitude $\kmskpc{15}$ in the northeast quadrant. In a completely independent method, a detailed model of the density and velocity structure in that part of the halo was generated. PV diagrams constructed from the data and the model were compared (both visually and by consideration of residual statistics); the results of this method confirmed the presence of a velocity gradient, with the same magnitude determined via the envelope tracing technique.

The results of this study are of interest with respect to recent observations of the neutral gas in the halo of NGC 891 \citep{foss05}. We have concluded that a vertical gradient in azimuthal velocity is present, of magnitude $\kmskpc{\approx 15-18}$, in the same area studied by that group. This value of the gradient is the same as has been determined for the \hi\ component (for the northeast quadrant alone) despite the very different radial distributions of the two components. 

Although extinction in the plane prohibits a determination of the major axis rotation curve from optical emission lines, evidence presented here suggests that the velocity gradient begins at approximately $z=0.6-1.6\,\mathrm{kpc}$. \citet{foss05} were unable to distinguish a corotating layer up to $z=1.3\,\mathrm{kpc}$ from the effects of beam smearing. We suggest that a thin corotating layer is the more likely interpretation.

In the southeast quadrant of the halo, the situation is quite different. Instead of a linear decline in rotation velocity with height, we find evidence for a constant rotation velocity ($\kms{\sim 175}$), significantly slower than the disk. Whether this, like the different EDIG morphology, is a consequence of the lower level of star formation activity in the southern disk is not yet clear, but the explanation for the discrepancy may have a significant impact on our understanding of the disk-halo interaction.

The ballistic model of \citet{cbr02} was unsuccessful in reproducing the halo kinematics of this galaxy. The velocity gradient in the model, driven mainly by the radially outward motion of clouds during their orbit through the halo, was found to be too shallow by more than an order of magnitude. To summarize, the envelope tracing method indicates $dv/dz=\kmskpc{17-18}$ for pointing H; the PV diagram modeling method indicates $dv/dz=\kmskpc{15}$ for pointing H, while in pointing L3 the data suggest a rapid decline in rotation speed with $z$, followed by a constant rotation velocity $\kms{\sim 175}$; in comparison, the ballistic model predicts a gradient of only about $dv/dz=\kmskpc{1-2}$.

Cuts through total intensity maps of the ballistic model and the \ha\ image of \citet{rkh90} were also compared. Because the ballistic model predicts larger vertical excursions where the galactic potential is weaker, the vertical extent of the halo grows with increasing radius. This characteristic density structure is not observed in the data. The velocity widths measured along the minor axis of NGC 891 do not show evidence for radial motion in the halo, but this result is uncertain.

Taken together, these results emphasize that the ballistic model, in its current form, is not sufficient to explain the dynamics of gaseous halos. Future models will need to provide a steeper vertical gradient in rotation velocity, while suppressing the tendency to produce halos of the ``flared'' appearance seen in the ballistic model. The hydrostatic and hydrodynamic models of \citet{bcfs06} and \citet{kmwsm05}, respectively, have proven successful in reproducing the gradient in NGC 891. We suggest that the next logical step may be to consider hybrid models consisting of quasi-ballistic particles orbiting within a (hydrodynamic or hydrostatic) gaseous halo, interacting with it via a drag force, for example.

The discrepancy between the data and the ballistic model is in the same sense as the results presented in Paper I for NGC 5775. We note, however, that the magnitude of the discrepancy is more pronounced in NGC 891 than in NGC 5775 (the ballistic model gradient was too low by only a factor of two in that case; see Paper I). The morphology of the EDIG in NGC 5775 is more filamentary than in NGC 891. Although \citet{rdwn04} observe vertical filamentary structures in the halo of NGC 891 using high-spatial resolution \emph{HST} images, and although pointing H covers two large, well-defined filaments, those features are far less pronounced relative to the underlying smooth EDIG component than is the case in NGC 5775. The differences in EDIG morphologies in NGC 5775 and NGC 891 may suggest that the disk-halo interaction in the former galaxy is closer to a pure galactic fountain, and thus the dynamics are more closely reproduced with ballistic motion. At present, we can only speculate on this possible relationship between the appearance of the halo gas and its dynamical evolution. It is also essential to recognize that NGC 5775 is experiencing an interaction with its companion NGC 5774, while NGC 891 appears to be far more isolated. Therefore, the differences in halo kinematics might alternatively be attributed to different levels of gas accretion. More observations of edge-ons are required to understand the relative importance of these effects. In a forthcoming paper, we present SparsePak observations of NGC 4302, completing a study of halo kinematics in a small sample of edge-ons with morphologically distinct EDIG emission. NGC 4302, which has the smoothest EDIG of the three galaxies, should have kinematics most different from the predictions of the ballistic model, if the appropriate class of disk-halo model is suggested by the appearance of the extraplanar gas. We note that NGC 4302 has a companion, NGC 4298; its possible interactions with that galaxy may also be an important factor.

\acknowledgments{We would like to thank F. Fraternali for providing the \hi\ rotation curves. We are greatly indebted to the support staff at NOAO/WIYN for their help in making the observing run successful. We thank an anonymous referee for comments which have improved the presentation of this paper. This material is based on work partially supported by the National Science Foundation under Grant No. AST 99-86113. R.~A.~B. acknowledges support from NASA ATP Grant NAG5-12128. M.~A.~B. acknowledges support by the National Science Foundation under Grant No. AST 03-07417, and thanks the University of Toronto for their hospitality.}

\bibliography{ms}

\clearpage
\begin{deluxetable}{cccccccc} 
\tabletypesize{\scriptsize}
\tablecaption{Observing Log}
\tablehead{\colhead{Pointing ID} & \colhead{RA\tablenotemark{a}} & \colhead{Decl.\tablenotemark{a}} & \colhead{Array PA} & \colhead{$R'$\tablenotemark{b}} & \colhead{$z$\tablenotemark{b}} & \colhead{Exp. Time\tablenotemark{c}} & \colhead{rms Noise\tablenotemark{d}}\\ \colhead{(see Fig. 1)} & \colhead{(J2000.0)} & \colhead{(J2000.0)} & \colhead{($\degr$)} & \colhead{($\arcsec$)} & \colhead{($\arcsec$)} & \colhead{(hr)} & \colhead{(erg s$^{-1}$ cm$^{-2}$ \AA$^{-1}$)}}
\tablecolumns{8}
\startdata
H & 02 22 42.84 & 42 22 24.10 & $-68$ & $-155$ to $-85$ & 30 to 98 (E) & 9.2 & $8.69 (8.06)\times 10^{-18}$\\
L1 & 02 22 38.39 & 42 20 31.44 & $-68$ & $-32$ to 38 & 27 to 95 (E) & 1.5 & $6.28 (6.15) \times 10^{-18}$\\
L2 & 02 22 27.71 & 42 21 19.24 & $+112$ & $-32$ to 38 & 33 to 101 (W) & 1.5 & $6.37 (6.12) \times 10^{-18}$\\
L3 & 02 22 33.94 & 42 18 38.78 & $-68$ & 91 to 161 & 23 to 91 (E) & 2.5 & $5.05 (4.80) \times 10^{-18}$\\
\enddata
\tablenotetext{a}{R.A. and Decl. of fiber 52 (the central non-``sky'' fiber in the SparsePak array) for each pointing.}
\tablenotetext{b}{Ranges of $R'$ and $z$ covered by each pointing of the fiber array. ``Sky'' fibers are not included in these ranges. $R'$ is positive on the south (receding) side. Letters E and W indicate that the pointing is on the east and west side of the disk, respectively. At $D=9.5\,\mathrm{Mpc}$, $22\arcsec=1\,\mathrm{kpc}$.}
\tablenotetext{c}{Total exposure time, which is the sum of individual exposures of about 30 minutes each.}
\tablenotetext{d}{The rms noise was measured in the continuum near the \ha\ line for each of the 82 fibers in every pointing. The tabulated values are the mean (median).}
\label{table:obslog}
\end{deluxetable} 

\clearpage

\end{document}